\documentclass[aps,prd,twocolumn,psfig,groupedaddress]{revtex4}
\newcommand{\fourvec}[1]{\mbox{\boldmath $\mathsf{#1}$}}

\begin{document}

\input psfig.sty


\title{Covariant Calculation of General Relativistic Effects 
in an Orbiting Gyroscope
Experiment}


\author{Clifford M. Will}
\email[]{cmw@wuphys.wustl.edu}
\homepage[]{wugrav.wustl.edu/People/CLIFF}
\affiliation{McDonnell Center for the Space Sciences, Department of
Physics, \\Washington University, St. Louis, Missouri 63130}


\date{\today}

\begin{abstract}
We carry out a covariant calculation of the measurable relativistic
effects in an orbiting gyroscope experiment.  The experiment,
currently known as Gravity Probe B, compares the spin directions of an
array of spinning gyroscopes with the optical axis of a telescope, all
housed in a spacecraft that rolls about the optical axis.  The
spacecraft is steered so that the telescope always points toward a
known guide star.   We calculate the variation in the spin directions
relative to readout loops rigidly fixed in the spacecraft, and express
the variations in terms of quantities that can be measured, to
sufficient accuracy, using an Earth-centered coordinate system.  The
measurable effects include the aberration of starlight, the geodetic
precession caused by space curvature, the frame-dragging effect
caused by the rotation of the Earth and the deflection of light by the
Sun.
\end{abstract}


\maketitle

\section{Introduction \label{intro}}
%
%
%
Gravity Probe B -- the ``gyroscope experiment'' -- is a NASA space
experiment designed to measure the general relativistic effect known as
the dragging of inertial frames.
The experiment will place into
Earth orbit a spacecraft containing four gyroscopes and a telescope (for
details of the experiment see \cite{nearzero,gpbweb}).  
The gyroscopes are 4 cm diameter fused quartz or single crystal
silicon
balls, machined to be as spherical as possible (to tolerances better
than one ppm), and coated with a thin film of niobium.  At
the low temperatures provided by a dewar of superfluid
Helium surrounding the gyroscope assembly, the niobium is
superconducting, and each spinning gyroscope has a London
magnetic moment parallel to its spin axis (stray magnetic fields, such
as those due to the Earth, will have been suppressed by many orders of
magnitude before the niobium goes superconducting).  The orientations
of the gyros' magnetic moments are measured via changes in the magnetic
flux through superconducting
current loops that encircle each gyro, and that are attached to the
gyroscope housing.  
The four gyros will
initially have their spins aligned parallel to the symmetry axis of
the spacecraft, which coincides with the optical axis of the telescope
mounted on the end of  the spacecraft.  In order to average out
numerous unwanted torques on the gyros, the spacecraft will rotate
about its symmetry axis at a rate of between one and 10 times per minute.  

According to general relativity, a perfect gyroscope in orbit about the Earth
will precess relative to distant inertial frames because of two
effects.  The first, and most important, is
the dragging of inertial frames, caused by the rotation
of the Earth, also called the Lense-Thirring effect.  Over a time of
one year, for a polar orbit of about 640 km altitude, this causes the
gyroscopes to precess in an East-West direction by around 42
milliarcseconds (mas).  The second is the geodetic precession, caused by the
curved spacetime around the Earth.  This effect results in an annual
precession in the North-South direction of about 6600 milliarcseconds.

The reference direction to the ``distant stars'' is provided by the
on-board telescope, which is to be trained on a star IM Pegasus 
(HR 8703) in our galaxy.
One important feature of this star is that it is also a strong radio
source, so that its direction and 
proper motion relative to the larger system of
astronomical reference frames can be measured accurately using Very
Long Baseline Radio Interferometry (VLBI).
In order to reduce telescope errors and biases, the spacecraft will be
``steered'' by attitude control forces to keep the centroid of the stellar image
centered on the telescope axis.  

However, because the spacecraft orbits the Earth and the Earth orbits
the Sun, the apparent direction to the star will vary in a periodic
manner because of
stellar aberration, which amounts to five arcseconds 
from the orbital motion, and to 20 arcseconds from the
annual motion of the Earth.  These are both substantially larger than the
frame-dragging effect being measured.  But, far from being an annoying 
ancillary effect getting in the way of testing relativity, the
aberration of starlight is in fact central to the success of the
experiment \cite{efs71}.  The reason is that the SQUID magnetometers that are
connected to the current loops surrounding the gyros do not measure
angle, they measure voltage, induced by the varying magnetic flux
threading the loops.  They must therefore be calibrated to provide a
scale factor that converts voltage to angle.  Because the orbit of the
spacecraft, the location of the star and the speed of light can all be
established to high precision, the aberration of starlight can be
predicted accurately.  It can also be separated from the relativistic
precessions because of its unique temporal variability over the 12+
months of the science phase of the mission.  
Through this means, the experiment can be
calibrated with an accuracy sufficient to achieve the desired overall accuracy.

In such a complex and delicate experiment, there are numerous sources
of error, including anomalous, non-relativistic torques on the gyros,
variations in electronics responses, errors in guide star position and
proper motion, effects of accumulated charges on the rotors from 
cosmic rays and other charged particles,
and so on.  Nevertheless, the team that is mounting this experiment --
Stanford University, Lockheed-Martin Space Systems, and NASA -- has
proposed that a measurement of the relativistic precessions at the
level of 0.4 mas/yr is feasible; this would provide a test of
relativistic frame dragging at the one percent level and of the
geodetic effect at the level of 6 parts in $10^5$.

The precession of gyroscopes according to general relativity was first
calculated independently by Pugh and by Schiff \cite{pugh,schiff1,schiff2}, 
who calculated the precession
relative to a given inertial coordinate system.  Wilkins
\cite{wilkins}
analysed the precession in more detail, attempting in particular to
ascertain how much of the geodetic effect could be regarded as Thomas
precession and how much was due to the curvature of space.  He
was also the first to attempt a more-or-less
coordinate-free calculation by
comparing the spin directions with the directions of incoming light
rays from distant sources.

Recent discussions of gyroscope precession generally fall into three groups.
One group provides
formal calculations of the precession of a gyroscope relative to a
local tetrad or congruence of geodesics, 
sometimes with applications to specific
spacetimes such as Schwarzschild, Kerr, or de Sitter 
\cite{rindler,tsoubelis,iyer,jantzen,pastora}, but seldom refers to
specific aspects of Gravity Probe-B.
Another group makes coordinate-dependent
calculations  of specific precession terms  (generally in the post-Newtonian
approximation), and discusses details of the various terms,
frequently with 
reference to Gravity Probe-B 
\cite{teyssandier,barkeroconnell,breakwell,adler}.  
This group also includes
unpublished calculations  by the GP-B team in the  course of
developing the data analysis procedures for the experiment.
A third group mixes calculations of precessions referred to a formal tetrad
(sometimes including comparisons with the direction to a distant star)
with applications to Gravity Probe-B; these include the standard
textbook presentations 
\cite{weinberg,mtw,tegp,soffel,brumberg}, as well as papers designed
in part to sort out the meaning of the various precession terms --
geodetic, frame-dragging and Thomas precession \cite{wilkins,ashby}.

To date, we know of no calculation that, on the one hand, 
takes a fully covariant
approach, and that on the other hand, incorporates critical 
operational aspects of the GP-B experiment, and
performs an end-to-end calculation in terms of observable quantities.
Such a covariant calculation is of interest for several
reasons.  It would verify
that this specific spacecraft
experiment is actually measuring the observable general
relativistic effects it claims to be measuring.
In addition, the effects of aberration are of order $v/c$ and are not
relativistic (they depend only on the finite propagation velocity of light),
however they are subject to relativistic corrections which are of
order $(v/c)^2$, comparable to the two general relativistic effects
being measured.  They should therefore be accounted for 
in a manner that does not depend on coordinates.

We have carried out such a calculation.  In Section II, we set up an
orthonormal basis of vectors tied to the rolling spacecraft and use
them to characterize the measurable components of the gyroscope spins as
read by current loops and SQUIDs on board.
Section III deals with the incoming light from the guide star and
incorporates the effects 
of spacecraft pointing on the spacecraft basis vectors.  In
Section IV we calculate the aberration and relativistic effects on the
observed spin components using the parametrized post-Newtonian framework
in an Earth-fixed reference frame.
Section IV discusses the various effects and estimates their
magnitudes, and
Section V gives concluding remarks.  We use the conventions of
\cite{tegp} and units in which $G=c=1$.

\section{Spacecraft Basis Vectors and Gyroscope Spins}

We consider a spacecraft with four-velocity $\fourvec{u}$, whose
orientation is described by
three spacelike basis vectors orthogonal to $\fourvec{u}$ 
that are rigidly tied to the spacecraft \cite{note1}.
The vector $\fourvec{\xi}$ points along the spacecraft symmetry axis,
while $\fourvec{\rho}$ and $\fourvec{\eta}$ are orthogonal to 
$\fourvec{\xi}$ and to each other.   These four vectors 
$\fourvec{e}_{\alpha}$ form an orthonormal basis, with 
$\fourvec{e}_{\alpha} \cdot \fourvec{e}_{\beta} = \eta_{\alpha\beta}$,
where $\eta_{\alpha\beta}$ is the Minkowski metric, valid in the local
inertial frame of the orbiting spacecraft.  The
spacecraft moves on a geodesic (we ignore atmospheric drag and translational
attitude control forces), with 
$\fourvec{\nabla}_{\fourvec{u}} \fourvec{u} = 0$.  Because
the spacecraft is rotating about its symmetry axis, it acts as a
gyroscope, so that its spin axis  $\fourvec{\xi}$ is parallel
transported, except for attitude-control forces that keep the
telescope pointed toward the guide star, so that 
$\fourvec{\nabla}_{\fourvec{u}} \fourvec{\xi}= \fourvec{t}$, where
$\fourvec{t}$ denotes the effect of those forces on $\fourvec{\xi}$
\cite{note2}. 
This, together with the orthonormality
of the basis and the geodesic equation, are sufficient to establish
that
\begin{eqnarray}
\fourvec{\nabla}_{\fourvec{u}} \fourvec{\eta} &=& \omega \fourvec{\rho} - 
	(\fourvec{\eta} \cdot \fourvec{t}) \fourvec{\xi} \,,
\nonumber\\
\fourvec{\nabla}_{\fourvec{u}} \fourvec{\rho} &=& -\omega \fourvec{\eta} -
	(\fourvec{\rho} \cdot \fourvec{t}) \fourvec{\xi} \,,
\label{eq1}
\end{eqnarray}
where
\begin{equation}
\omega \equiv \fourvec{\rho} \cdot \fourvec{\nabla}_{\fourvec{u}}\fourvec{\eta} 
	= - \fourvec{\eta} \cdot \fourvec{\nabla}_{\fourvec{u}}\fourvec{\rho}
\label{eq2}
\end{equation}
is the locally measured roll angular frequency of the spacecraft.
Defining the spacecraft roll phase $\theta \equiv \int \omega d\tau$,
where $\tau$ is spacecraft proper time, we define ``fixed'' roll reference
axes which do not rotate with the spacecraft:
\begin{eqnarray}
\fourvec{\eta}_0 \equiv \fourvec{\eta} \cos \theta - \fourvec{\rho}
\sin \theta \,,
\nonumber \\
\fourvec{\rho}_0 \equiv \fourvec{\eta} \sin \theta + \fourvec{\rho}
\cos \theta \,.
\label{eq3}
\end{eqnarray}

The initial phase is chosen so that one of the vectors ($\fourvec{\eta}_0$)
points
orthogonal to the plane of the polar orbit, or in the East-West (EW) direction, while
the other ($\fourvec{\rho}_0$) 
lies in the orbital plane, or the North-South (NS) direction (see Fig.
1).  
The orbital plane is
chosen so that the guide star, and hence the vector $\fourvec{\xi}$,
lie in the plane.  (The guide star HR 8703 is at declination
$16.84^{\rm o}$ and right ascension $343.26^{\rm o}$.)
Then one can show that these roll-fixed vectors evolve according to
\begin{eqnarray}
\fourvec{\nabla}_{\fourvec{u}} \fourvec{\eta}_0 &=& -
        (\fourvec{\eta}_0 \cdot \fourvec{t}) \fourvec{\xi} \,,
\nonumber\\
\fourvec{\nabla}_{\fourvec{u}} \fourvec{\rho}_0 &=& -
        (\fourvec{\rho}_0 \cdot \fourvec{t}) \fourvec{\xi} \,,
\nonumber\\
\fourvec{\nabla}_{\fourvec{u}} \fourvec{\xi} &=& \fourvec{t} \,,
\label{eq4}
\end{eqnarray}
where we have used the fact that $\fourvec{t} \cdot \fourvec{u} =0$.  

\begin{figure}
\leavevmode
\psfig{figure=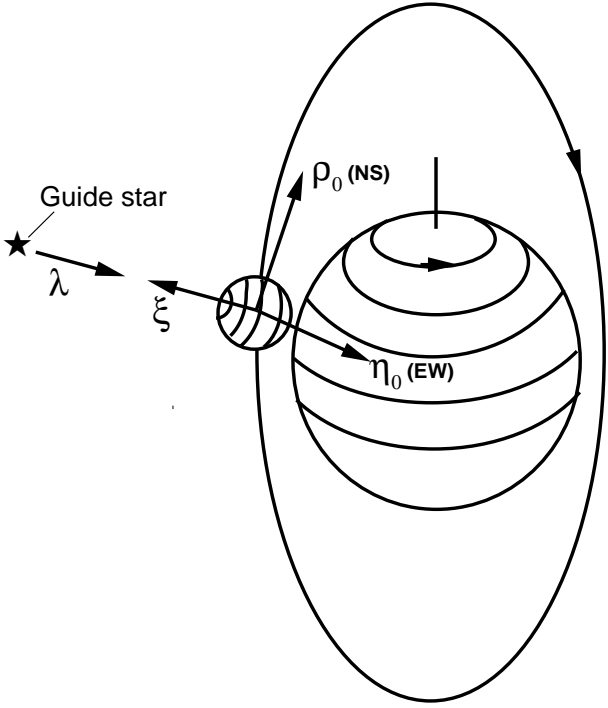,width=8cm}
\caption{\label{Fig1} Basis four-vectors for Gravity Probe-B}
\end{figure}

Each gyroscope is characterized by a spin four-vector $\fourvec{S}$ 
that is purely spatial in the spacecraft frame 
($\fourvec{S} \cdot \fourvec{u} =0$); this amounts to a specific choice of its
center of mass.  
We assume that gyro torques are sufficiently small that we
can treat the gyros as perfect, so that $\fourvec{S}$ is parallel transported 
($\fourvec{\nabla}_{\fourvec{u}} {\fourvec{S}} = 0$).
Then the rate of change of each component of $\fourvec{S}$ on the 
roll-fixed basis is given by $d(\fourvec{S} \cdot \fourvec{e}_\alpha)/d\tau =
\fourvec{S} \cdot \fourvec{\nabla}_{\fourvec{u}} \fourvec{e}_\alpha$; 
combining with Eqs.
(\ref{eq4}) we obtain the equation of motion for the spin
components,
\begin{eqnarray}
\dot S_{\eta_0} &=& - t_{\eta_0} S_\xi \,,
\nonumber \\
\dot S_{\rho_0} &=& - t_{\rho_0} S_\xi \,,
\nonumber \\
\dot S_\xi &=& t_{\eta_0} S_{\eta_0} + t_{\rho_0} S_{\rho_0} \,,
\label{eq5}
\end{eqnarray}
where $S_{\eta_0} = \fourvec{S} \cdot \fourvec{\eta}_0$
$S_{\rho_0} = \fourvec{S} \cdot \fourvec{\rho}_0$, 
and $S_\xi = \fourvec{S} \cdot \fourvec{\xi}$, 
with parallel definitions for $t_{\eta_0}$, etc., and
where overdot denotes $d/d\tau$.   In the absence of torques on the
spacecraft ($\fourvec{t}=0$), $\dot S_{\eta_0} = \dot S_{\rho_0} =\dot S_\xi = 0$, as
expected, since the gyro spins 
and spacecraft basis vectors will undergo identical
transports through spacetime.  

\section{Telescope pointing}

The spacecraft is oriented so that the optical axis of the telescope is
aligned parallel to the direction of an incoming photon from the guide
star.  Let $\fourvec{\lambda}$ be the null tangent vector to the incoming
photon 
satisfying the geodesic equation 
$\fourvec{\nabla}_{\fourvec{\lambda}} \fourvec{\lambda} = 0$
with 
$\fourvec{\lambda} \cdot \fourvec{\lambda}=0$. 
In the spacecraft frame, the direction of the incoming photon is given
by a unit normalized spatial vector which is the projection of 
$\fourvec{\lambda}$ onto the spacecraft basis, namely
\begin{equation}
{\fourvec{\hat \lambda}} = {{\fourvec{\lambda}} \over {(-\fourvec{u}\cdot
\fourvec{\lambda})}} - \fourvec{u} \,.
\label{nullvector}
\end{equation}
Then, using Eqs. (\ref{eq4})
the equations of motion for the roll-fixed components of
$\fourvec{\hat \lambda}$ are given by
\begin{eqnarray}
(d/d\tau) {\hat \lambda}_{\eta_0} &=& 
\epsilon^{1/2} (\fourvec{\nabla}_{\fourvec{u}} \fourvec{\hat \lambda}) 
\cdot \fourvec{\eta}_0 - t_{\eta_0} {\hat \lambda}_\xi \,,
\nonumber \\
(d/d\tau) {\hat \lambda}_{\rho_0} &=& 
\epsilon^{1/2} (\fourvec{\nabla}_{\fourvec{u}} \fourvec{\hat \lambda}) 
\cdot \fourvec{\rho}_0 - t_{\rho_0} {\hat \lambda}_\xi \,,
\nonumber \\
(d/d\tau) {\hat \lambda}_\xi &=& 
\epsilon^{1/2} (\fourvec{\nabla}_{\fourvec{u}} \fourvec{\hat \lambda}) 
\cdot \fourvec{\xi} + t_{\eta_0} {\hat \lambda}_{\eta_0} 
+ t_{\rho_0} {\hat \lambda}_{\rho_0}
\,,
\label{eq7}
\end{eqnarray}
where we have introduced the small parameter $\epsilon^{1/2} \sim (v/c)$
to characterize the leading effects on 
$\fourvec{\nabla}_{\fourvec{u}} \fourvec{\hat \lambda}$,
namely aberration.

In the operation of the experiment, attitude control forces orient the
spacecraft so that the telescope always points toward the apparent
location of the guide star, i.e. so that 
$\fourvec{\hat \lambda} = - \fourvec{\xi}$, 
or so that 
${\fourvec{\hat \lambda}} \cdot \fourvec{\rho}_0 
= \fourvec{\hat \lambda} \cdot \fourvec{\eta}_0 = 0$.  
This will be only approximately true because
of pointing errors and occasional deliberate ``dithering'' of the
spacecraft for operational and other reasons; these errors are
expected to be on the level of a few to tens of mas.  
We introduce the small parameter
$\delta \sim 10^{-7}$ to characterize these pointing errors.  
Then to drive ${\hat \lambda}_{\rho_0}$ and ${\hat \lambda}_{\eta_0}$ 
and their time
derivatives to zero,
we require attitude control torques given, from Eqs. (\ref{eq7}), by
\begin{eqnarray}
t_{\eta_0} &=& \epsilon^{1/2} {\hat \lambda}_\xi^{-1} 
(\fourvec{\nabla}_{\fourvec{u}} \fourvec{\hat \lambda}) \cdot
{\fourvec\eta}_0 + \delta (\Delta t_{\eta_0}) \,,
\nonumber \\
t_{\rho_0} &=& \epsilon^{1/2} {\hat \lambda}_\xi^{-1} 
(\fourvec{\nabla}_{\fourvec{u}} \fourvec{\hat \lambda}) \cdot
{\fourvec\rho}_0 + \delta (\Delta t_{\rho_0}) \,, 
\label{eq8}
\end{eqnarray}
where $\Delta t_{\eta_0}$ and $\Delta t_{\rho_0}$ are the roll-fixed
components of the order-$\delta$ residual ``pointing error'' and dithering
torques.  Substituting Eqs. (\ref{eq8})
back into Eqs. (\ref{eq7}), and using the fact that 
${\fourvec{\hat \lambda}}$ 
is a normalized vector, we find that 
$(d/d\tau) {\hat \lambda}_{\eta_0} = - \delta (\Delta t_{\eta_0}) 
	{\hat \lambda}_\xi$, 
$(d/d\tau) {\hat \lambda}_{\rho_0} = - \delta (\Delta t_{\rho_0}) 
	{\hat \lambda}_\xi$, and 
$(d/d\tau) {\hat \lambda}_\xi =  
\delta (\Delta t_{\eta_0} {\hat \lambda}_{\eta_0} 
+ \Delta t_{\rho_0} {\hat \lambda}_{\rho_0} )$.
From these equations we obtain 
\begin{equation}
{\hat \lambda}_{\eta_0}(\tau) = O(\delta) \,,
{\hat \lambda}_{\rho_0}(\tau) = O(\delta) \,,    
{\hat \lambda}_\xi(\tau) = -1 + O(\delta^2) \,.
\label{lambdasol}
\end{equation}

Then substituting Eqs. (\ref{eq8}) into (\ref{eq5}), we obtain
\begin{eqnarray}
{\dot S}_{\eta_0} &=& -\epsilon^{1/2} (S_\xi/{\hat \lambda}_\xi) 
(\fourvec{\nabla}_{\fourvec{u}} \fourvec{\hat \lambda}) 
\cdot \fourvec{\eta}_0 - \delta (\Delta t_{\eta_0}) 
S_\xi \,,
\nonumber \\
{\dot S}_{\rho_0} &=& -\epsilon^{1/2} (S_\xi/{\hat \lambda}_\xi) 
(\fourvec{\nabla}_{\fourvec{u}} \fourvec{\hat \lambda}) 
\cdot \fourvec{\rho}_0 - \delta (\Delta t_{\rho_0})
S_\xi \,,
\nonumber \\
{\dot S}_\xi &=& \epsilon^{1/2} [S_{\eta_0} 
(\fourvec{\nabla}_{\fourvec{u}} \fourvec{\hat \lambda}) 
\cdot \fourvec{\eta}_0 + S_{\rho_0} 
(\fourvec{\nabla}_{\fourvec{u}} \fourvec{\hat \lambda}) 
\cdot \fourvec{\rho}_0 ] {\hat \lambda}_\xi^{-1}
\nonumber \\
&& + \delta (\Delta t_{\eta_0} S_{\eta_0} + \Delta t_{\rho_0} S_{\rho_0} ) \,.
\label{eq9}
\end{eqnarray}
But since $S_{\eta_0} \sim S_{\rho_0} \sim \epsilon^{1/2} + \delta$, we have
${\dot S}_\xi \sim \epsilon + \epsilon^{1/2}\delta + \delta^2 +
\dots$, and so, choosing the vector $\fourvec{S}$ to be normalized to
unity (under parallel transport it has constant norm), 
we can write $S_\xi = 1 + O(\epsilon, \epsilon^{1/2}\delta,
\delta^2)$.  Substituting into Eq. (\ref{eq9}), we obtain our final
equation for the evolution of the gyro spin components relative to the
roll-fixed spacecraft axes, 
\begin{eqnarray}
{\dot S}_{\eta_0} &=& \epsilon^{1/2}  
(\fourvec{\nabla}_{\fourvec{u}} \fourvec{\hat \lambda}) 
\cdot \fourvec{\eta}_0 - \delta (\Delta t_{\eta_0})
\nonumber \\
&&
+ O(\epsilon^{3/2}, \epsilon \delta, \epsilon^{1/2} \delta^2) \,,
\nonumber \\
{\dot S}_{\rho_0} &=& \epsilon^{1/2}  
(\fourvec{\nabla}_{\fourvec{u}} \fourvec{\hat \lambda}) 
\cdot \fourvec{\rho}_0 - \delta (\Delta t_{\rho_0})
\nonumber \\
&&
+ O(\epsilon^{3/2}, \epsilon \delta, \epsilon^{1/2} \delta^2) \,.
\label{eq10}
\end{eqnarray}
These physically measurable 
components $S_{\eta_0}$ and $S_{\rho_0}$
translate directly into outputs of the SQUID
magnetometers.  The expressions are generally covariant, and thus may
be evaluated in any chosen coordinate system.

Interestingly, in this picture all the relativistic effects 
arise mathematically
from the covariant derivative of the incoming photon world line
along the spacecraft four-velocity, and not in the precessions of the
gyroscope spins.  This is because we have chosen the rolling
spacecraft itself to define our basis; this was a natural choice because all
the measuring instruments (current loops, SQUIDs) 
are rigidly fixed to the spacecraft.  But
since a rolling spacecraft is itself a gyroscope, it and the superconducting
gyros precess together in the spacetime around the Earth.  Hence, in this basis, the relativistic effects appear to
originate elsewhere.  The superconducting gyros are still crucial, of
course, because, by design, they are closer to being perfect gyros,
and thus are less affected by anomalous
torques.  Here the spacecraft plays a role similar to that of an
oscillator clock that is slaved to an ultraprecise atomic
standard (such as a hydrogen maser, or atom fountain clock); the
oscillator may drift or be deliberately steered from time to time, but
its phase can always be linked precisely back to the underlying
standard.  Similarly, while the spacecraft ``gyro'' is steered to follow
the guide star as the star's apparent location varies because of aberration
and relativistic effects, the spacecraft axis 
direction is precisely linked to the
superconducting gyros via the output of the SQUIDS.  

We now turn to the relativistic and aberration effects that are being
measured.

\section{Post-Newtonian calculation of measurable relativistic effects}

We evaluate the quantity 
$\fourvec{\nabla}_{\fourvec{u}} \fourvec{\hat \lambda}$
in the post-Newtonian approximation using
the parametrized post-Newtonian (PPN) framework \cite{tegp}.  The metric
of spacetime in the solar-system rest-frame is taken to have the form
\begin{eqnarray}
g_{00} &=& -1 + 2U + O(\epsilon^2)\,,
\nonumber \\
g_{0i} &=& -(2 + 2\gamma + \frac{1}{2} \alpha_1 ) V_i +
O(\epsilon^{5/2})\,,
\nonumber \\
g_{ij} &=& (1 + 2\gamma U)\delta_{ij}+ O(\epsilon^{2}) \,,
\label{metric}
\end{eqnarray}
where $\gamma$ and $\alpha_1$ are PPN parameters whose values
depend on the theory of gravity, and 
where we have restricted attention to theories of gravity with
suitable global conservation laws, and have ignored
``preferred-frame'' terms in the metric (see \cite{tegp}, \S 9.1 for
discussion of these effects).  We have used a gauge analogous to
harmonic gauge, instead of the conventional PPN gauge (see \cite{tegp}
\S 4.2 for discussion).  For an orbiting, rotating, nearly
spherical Earth, and a static, spherically symmetric Sun,
the potentials 
$U$ and ${\bf V}$ have the form
\begin{eqnarray}
U &=& M_\odot/r + M_\oplus/R\,, 
\nonumber \\
{\bf V} &=& M_\oplus {\bf v}_\oplus/R 
 - \frac{1}{2} {\bf X} \times {\bf J_\oplus} /R^3 \,,
\label{UV}
\end{eqnarray}
where $M_\odot$ is the mass of the Sun,
$M_\oplus$, ${\bf v}_\oplus$ and ${\bf J}_\oplus$ are the 
mass, velocity and angular momentum of the Earth; 
${\bf X}={\bf x} - {\bf x}_\oplus$ and  $R=|{\bf X}|$ denote the 
satellite-Earth vector and distance, while $\bf x$ and $r = |{\bf x}|$ 
denote the satellite-Sun
vector and distance.  In fact, the quadrupole moment of the Earth has
a measurable effect on the geodetic precession and must be included;
as this has been treated thoroughly elsewhere
\cite{wilkins,barker70,breakwell,adler} we will ignore it here.

From the geodesic equation for photons, $d \lambda^i /dt +
(\Gamma^i_{\mu\nu}  - \Gamma^0_{\mu\nu} \lambda^i) \lambda^\mu \lambda^\nu
=0$, where $\lambda^\mu = dx^\mu/dt$, we obtain the solution
(see \cite{tegp} \S 7.1 for details):
\begin{eqnarray}
\lambda^0 &=& 1 \,,
\nonumber \\
\lambda^i &=& {n}^i [1 - (1+\gamma)U ] + {\cal D}^i +
O(\epsilon^{3/2}) \,,
\label{lambda}
\end{eqnarray}
where ${n}^i$ is a Cartesian unit vector describing the initial
direction of the photon far from the Earth in the global PPN
coordinate system, and ${\cal D}^i$ represents the deflection of the
ray, obtained by solving the equation
\begin{equation}
\left ( {{d {\cal D}^i} \over dt} \right )_{{\rm along}\, n^i}
	= (1+\gamma) [ \nabla^i U - n^i ({\bf n} \cdot \nabla U) ] \,.
\label{calD}
\end{equation}
The unit vector ${\bf n}$ points from the source toward the spacecraft,
and hence may be written ${\bf n} = {\bf n}_0 + {\bf x} \cdot \nabla
{\bf n}_0 + \dots$, where $ {\bf n}_0$ is a constant unit vector toward the
solar-system barycenter and where $\nabla^i n_0^j = (\delta^{ij} -
n_0^i n_0^j )/D$, where $D$ is the distance from the solar-system
barycenter to the guide star.   The correction term accounts for the
parallax of the guide star; for HR 8703 it amounts to about 10 mas
$\sim 3 \times 10^{-7}$.   Because it is so small, we will treat it as
{\it effectively} a pointing error term, of order $\delta$.

Letting the spacecraft have ordinary velocity $v^i = u^i/u^0$, where
$u^0 = 1 + v^2/2 +U + O(\epsilon^2)$, we find that the components of
${\fourvec{\hat \lambda}}$ are given by
\begin{eqnarray}
{\hat \lambda}^0 &=& {\bf n} \cdot {\bf v} - [v^2 - ({\bf n} \cdot
	{\bf v})^2 ] + O(\epsilon^{3/2}) \,,
\nonumber \\
{\hat \lambda}^i &=& {n}^i - [v^i - n^i ({\bf n} \cdot {\bf v})] + 
	n^i [({\bf n} \cdot {\bf v})^2 - v^2 /2 -\gamma U ] 
\nonumber \\
&&
+ {\cal D}^i +
	O(\epsilon^{3/2}) \,.
\label{lambdahat}
\end{eqnarray}

To calculate  the covariant quantities
$(\fourvec{\nabla}_{\fourvec{u}} \fourvec{\hat \lambda})
\cdot \fourvec{\eta}_0$ and 
$(\fourvec{\nabla}_{\fourvec{u}} \fourvec{\hat \lambda})
\cdot \fourvec{\rho}_0$ in Eqs. (\ref{eq10}), it suffices to  
evaluate them in the PPN coordinate system.
We first evaluate  
$(\fourvec{\nabla}_{\fourvec{u}} \fourvec{\hat \lambda})^\alpha = 
u^\beta {{\hat \lambda}^\alpha}_{; \beta}$.  Calculating the
Christoffel symbols from the metric (\ref{metric}) to the needed order,
and using the fact that $dU/dt=\partial U/\partial t + {\bf v} \cdot
\nabla U$, and, by virtue of the equation of motion $d{\bf v}/dt
= \nabla U$, that $({\bf v} \cdot {\bf n})\nabla U = 1/2 \{ d[{\bf v}
({\bf v} \cdot {\bf n})]/dt + {\bf n} \times (\nabla U \times {\bf
v})\}$, we find
\begin{eqnarray}
(\fourvec{\nabla}_{\fourvec{u}} \fourvec{\hat \lambda})^0 &=& O(v\nabla U) \,,
\nonumber \\
(\fourvec{\nabla}_{\fourvec{u}} \fourvec{\hat \lambda})^i &=& {d \over dt}
	\biggl [({\bf x} \cdot \nabla) n_0^i -v^i + n^i ({\bf n} \cdot {\bf v}) 
\nonumber \\
&&
	+ n^i \{({\bf n} \cdot {\bf v})^2 - \frac{1}{2} v^2 \}
	- \frac{1}{2} v^i ({\bf n} \cdot {\bf v}) + {\cal D}^i \biggr ]
\nonumber \\
	&& + \frac{1}{2} (2\gamma +1) [{\bf n} \times ({\bf v} \times 
	\nabla U)]^i 
	\nonumber \\
	&&+  (1+\gamma+ \frac{1}{4} \alpha_1 ) 
	[{\bf n} \times (\nabla \times {\bf V})]^i 
\nonumber \\
        &&+ O(U\nabla U) \,,
\label{nablalambda}
\end{eqnarray}
where,
to the order needed, we can convert from spacecraft proper time 
$\tau$ to
PPN coordinate time $t$.

We also need to determine the time dependence of
$\fourvec{\eta}_0$ and $\fourvec{\rho}_0$; even though they are roll-fixed
vectors, they will vary slightly because of pointing of the spacecraft to
maintain alignment with the guide star.  However, because 
$\fourvec{\nabla}_{\fourvec{u}} \fourvec{\hat \lambda}$ is already of
$O(\epsilon^{1/2})$, we only need to evaluate $\fourvec{\eta}_0$ and
$\fourvec{\rho}_0$ to $O(\epsilon^{1/2})$, which means accounting only
for the effects of aberration-induced torques.  
Substituting Eqs. (\ref{eq8}) into (\ref{eq4}), and using the fact
that $\fourvec{\xi}$, $\fourvec{\rho}_0$, $\fourvec{\eta}_0$, and
$\fourvec{u}$ form an orthonormal tetrad,
we obtain
\begin{eqnarray}
\fourvec{\nabla}_{\fourvec{u}} \fourvec{\eta}_0 &=& \epsilon^{1/2}
 \fourvec{\nabla}_{\fourvec{u}} \fourvec{\hat \lambda} \cdot
\fourvec{\eta}_0 \fourvec{\xi} + O(\delta) 
\,,
\nonumber \\
\fourvec{\nabla}_{\fourvec{u}} \fourvec{\rho}_0 &=& \epsilon^{1/2}
 \fourvec{\nabla}_{\fourvec{u}} \fourvec{\hat \lambda} \cdot
\fourvec{\rho}_0 \fourvec{\xi} + O(\delta) 
\,,
\nonumber \\
\fourvec{\nabla}_{\fourvec{u}} \fourvec{\xi} &=&  \epsilon^{1/2}
 (-\fourvec{\nabla}_{\fourvec{u}} \fourvec{\hat \lambda}
 +\fourvec{\nabla}_{\fourvec{u}} \fourvec{\hat \lambda} \cdot 
 \fourvec{u} \fourvec{u}
 + \fourvec{\nabla}_{\fourvec{u}} \fourvec{\hat \lambda} \cdot 
 \fourvec{\xi} \fourvec{\xi} ) 
 \nonumber \\
 &&+ O(\delta) \,.
\label{nablaeta}
\end{eqnarray}
We use the fact that, for each of these three vectors,
$(\fourvec{\nabla}_{\fourvec{u}} \fourvec{e}_{(\alpha)})^\beta =
d{e}_{(\alpha)}^\beta /d\tau + 
\Gamma^\beta_{\gamma\delta} e_{(\alpha)}^\gamma u^\delta$, substitute
the Christoffel symbols and Eqs. (\ref{nablalambda}), and
integrate, to obtain
\begin{eqnarray}
(\eta_0)^0 (t) &=& {\bf v} \cdot {\bf {\bar \eta}}_0  + O(\epsilon,
\delta)\,, 
\nonumber \\
(\eta_0)^i (t) &=& ({\bar \eta}_0)^i - {\bf v}\cdot {\bf {\bar \eta}}_0
   {\bar \xi}^i + O(\epsilon, \delta) \,,
\nonumber \\
(\rho_0)^0 (t) &=& {\bf v} \cdot {\bf {\bar \rho}}_0  + O(\epsilon,
\delta)\,, 
\nonumber \\
(\rho_0)^i (t) &=& ({\bar \rho}_0)^i - {\bf v}\cdot {\bf {\bar \rho}}_0
  {\bar \xi}^i + O(\epsilon, \delta) \,,
\nonumber \\
\xi^0 (t) &=& {\bf v} \cdot {\bf {\bar \xi}} + O(\epsilon,
\delta)\,, 
\nonumber \\
\xi^i (t) &=& {\bar \xi}^i + v^i - {\bar \xi}^i {\bf v} \cdot {\bf
\bar \xi} + O(\epsilon, \delta) \,,
\label{vecsolution}
\end{eqnarray}
where ${\bf {\bar \eta}}_0$, ${\bf {\bar \rho}}_0$ and ${\bf {\bar \xi}}$
are spatial 3-vectors in the PPN frame that describe the  
orientation of the spacecraft at a chosen initial moment of time.  They
are orthonormal in the Cartesian sense to $O(\epsilon)$.   We
have used the fact, obtainable from Eq. (\ref{lambdasol}), that ${\bf n} = -
{\bf {\bar \xi}} + O(\epsilon, \delta)$, ${\bf n} \cdot {\bf {\bar \eta}}_0 =
O(\epsilon, \delta)$ and  ${\bf n} \cdot {\bf {\bar \rho}}_0 =
O(\epsilon, \delta)$. 

Substituting  Eqs. (\ref{nablalambda}) and (\ref{vecsolution}) 
into (\ref{eq10}), converting from $d/d\tau$ to $d/dt$ using the
property that $\tau = t [1+O(\epsilon)]$, 
and integrating with respect to $t$, we obtain,
finally
\begin{eqnarray}
S_{\eta_0} &=& S_{\eta_0} (0) - \epsilon^{1/2}{\bf v} \cdot {\bf {\bar \eta}}_0
\nonumber \\
	&&
	+ \epsilon \left ( \frac{1}{2}{\bf v} \cdot {\bf {\bar \eta}}_0
		{\bf v} \cdot {\bf {\bar \xi}}  
	+ {\bf {\cal H}} \cdot {\bf {\bar \eta}}_0
	+ {\bf {\cal D}} \cdot {\bf {\bar \eta}}_0 \right )
\nonumber \\
	&&
	+ \delta (D^{-1} {\bf x} \cdot {\bf {\bar \eta}}_0
        - {\cal T}_{\eta_0} )
\nonumber \\
        &&
	+ O(\epsilon^{3/2},\epsilon \delta, \epsilon^{1/2} \delta^2)
\,,
\nonumber \\
S_{\rho_0} &=& S_{\rho_0} (0) - \epsilon^{1/2}{\bf v} \cdot {\bf \bar
\rho}_0
\nonumber \\
        &&
        + \epsilon \left ( \frac{1}{2}{\bf v} \cdot {\bf {\bar \rho}}_0
                {\bf v} \cdot {\bf {\bar \xi}}  
        + {\bf {\cal H}} \cdot {\bf {\bar \rho}}_0
        + {\bf {\cal D}} \cdot {\bf {\bar \rho}}_0 \right )
\nonumber \\
        &&
	+ \delta ( D^{-1} {\bf x} \cdot {\bf {\bar \rho}}_0
        - {\cal T}_{\rho_0} )
\nonumber \\
        &&
        + O(\epsilon^{3/2},\epsilon \delta, \epsilon^{1/2} \delta^2)
\,.
\label{finalanswer}
\end{eqnarray}
The first term in Eqs. (\ref{finalanswer}) is the initial misalignment
of the spin with the spacecraft axis, which is expected to be smaller
than one arcsecond.  The next term, of order $\epsilon^{1/2}$, and
the first of the $O(\epsilon)$ terms, are the aberration
term and its relativistic correction.  The term involving
${\bf {\cal H}}$ is the integrated relativistic precession projected
onto ${\bf {\bar \eta}}_0$ and ${\bf {\bar \rho}}_0$, given by 
${\bf {\cal H}} = \int^t (d{\bf {\cal H}}/dt) dt$, where $ d{\bf {\cal
H}}/dt$ is the precession rate given by
\begin{equation}
{d{\bf {\cal H}} \over dt} = -  
\frac{1}{2} (2\gamma +1) {\bf {\bar \xi}} \times ({\bf v} \times
        \nabla U) - (1+\gamma +\frac{1}{4} \alpha_1 ) 
	{\bf {\bar \xi}} \times (\nabla \times {\bf V}) \,.
\end{equation}
The terms ${\bf {\cal D}} \cdot {\bf {\bar \eta}}_0$ 
and ${\bf {\cal D}} \cdot {\bf {\bar \rho}}_0$ in Eqs. (\ref{finalanswer})
denote the contribution of the deflection of
light.  The $O(\delta)$ terms ${\cal T}_{\eta_0}$ and ${\cal T}_{\rho_0}$ 
denote the integrated angular offset
caused by pointing errors, while the terms 
$D^{-1} {\bf x} \cdot {\bf {\bar \eta}}_0$
and $D^{-1} {\bf x} \cdot {\bf {\bar \rho}}_0$ denote the effects of parallax.  
The spatial vectors defined in
Eqs. (\ref{finalanswer}) are referred to the solar-system-fixed PPN
coordinate system.  However, to the order needed, we can write ${\bf
v} = {\bf v}_\oplus + {\bf v}_s + O(\epsilon^{3/2})$, where ${\bf
v}_\oplus$ is the Earth's orbital velocity relative to the Sun, 
and ${\bf v}_s$ is the
spacecraft's velocity relative to the Earth.  Furthermore, the
components of the various 3-vectors in the solar-system basis are
equal to the respective components in a non-rotating Earth-comoving
basis, apart from corrections of order $\epsilon$; hence to the order
needed in Eqs. (\ref{finalanswer}) we can treat all spatial vectors as
defined in the geocentric basis.

\section{Discussion of Relativistic Effects}

The first non-constant term in Eqs. (\ref{finalanswer}) is the aberration of
starlight, consisting of both an annual aberration with amplitude
of  ${\bf v}_\oplus \sim 20$ arcseconds, 
and an orbital aberration with amplitude ${\bf v}_s \sim 5$
arcseconds.  But because 
the position of the
guide star can be determined  from VLBI and the orbit of the
earth and spacecraft can be determined
from the ephemeris and GPS to high accuracy, these terms may be
predicted to an accuracy better than that needed to measure the
relativistic effects.  Furthermore, their variation with time over a
year can be separated from the variation of the relativistic terms,
hence the aberration can be used to determine the ``scale factor''
that relates the output of the SQUIDs (the actual data)
to the desired angles $S_{\eta_0}$ and
$S_{\rho_0}$.   The second term in Eqs. (\ref{finalanswer}) is the
relativistic correction to aberration, whose orbital part dominates,
with amplitude 
$\frac{1}{2} v_\oplus^2
\sin \beta \sim 0.4$ mas ($\beta$ is the ecliptic latitude of the
guide star), which is comparable to the
accuracy goal of the experiment; this term must be included in the
data analysis \cite{stumpff}.  
Terms of order $v_\oplus v_s$ and $v_s^2$ and terms 
cubic in $v$ can be ignored.  

For the relativistic precession terms, $d {\bf {\cal H}}/dt$ can
be separated into various contributions using Eqs. (\ref{UV}):  
\begin{eqnarray}
{{d{{\bf \cal H}}} \over dt} &=&
\left \{ \frac{1}{4} (2 + \alpha_1 ) 
{M_\oplus \over R^3} {\bf v}_\oplus \times {\bf X} 
- \frac{1}{2} (2\gamma +1) {M_\odot \over r^3} {\bf v}_s \times {\bf x}
\right .
\nonumber \\
&&
\left .
-  \frac{1}{2} (2\gamma +1) {M_\odot \over
r^3} {\bf v}_\oplus \times {\bf x} -
\frac{1}{2} (2\gamma +1){M_\oplus \over R^3} {\bf v}_s \times {\bf X}
\right.
\nonumber \\
&&
\left .
- \frac{1}{2} (\gamma + 1 + \frac{1}{4} \alpha_1 ) 
\left [ {{{\bf J}_\oplus - 3 {\bf N} {\bf J}_\oplus \cdot  {\bf N}} 
	\over R^3 } \right ] \right \}
	\times {\bf {\bar \xi}} 
\,,
\label{Hdot}
\end{eqnarray}
where ${\bf N} = {\bf X}/R$.  Since $\cal H$ and ${\bf \bar \xi}$ both
refer to the direction of the spins, this equation has the general
form $d{\bf S} /dt = {\bf \Omega} \times {\bf S}$, so that the
quantity in braces in Eq. (\ref{Hdot}) can be regarded as the
precession rate vector.

The first two terms produce periodic contributions to the
signal at the frequencies $\omega_s \pm \omega_\oplus$, where
$\omega_s$ and $\omega_\oplus \ll \omega_s$ are the orbital angular 
frequencies  of the spacecraft and Earth,
respectively.  Integrated over time, 
these contribute periodic terms at approximately the orbital frequency
with amplitudes of
$v_\oplus M_\oplus / 2R^2 \omega_s \sim v_\oplus v_s/2 \sim 0.3$ mas 
and 
$v_s M_\odot /r^2 \omega_s \sim v_\oplus^2 (R/r) \sim 10^{-4}$ mas,
respectively.  Averaged over time, these are negligible.

The third term may be evaluated to sufficient accuracy at the center
of the Earth rather than at the satellite; treating the Earth's orbit
as circular yields a signal that grows at the constant
rate of $(3/2)v_\oplus^2
\omega_\oplus ({\bf h} \times {\bf {\bar \xi}} ) \sim 19$ mas/yr 
in a direction perpendicular to ${\bf h}$, the normal to the ecliptic
plane \cite{wilkins,barkeroconnell2}.  
This term is the solar geodetic effect (sometimes called the de Sitter
precession); it is responsible for
an analogous precession 
in the lunar orbit that has been measured to around 0.7 \% using Lunar
laser ranging \cite{williams}.  

The fourth term is the main geodetic effect, at the constant rate of
$(3/2)v_s^2
\omega_s ({\bf h}^\prime \times {\bf {\bar \xi}} ) \sim 6600$
mas/yr, in a direction perpendicular to ${\bf h}^\prime$, the normal to the
orbital
plane.
Note that if ${\bf {\bar \eta}}_0$ is chosen to be the EW basis vector, which
is perpendicular to the polar orbit, then the main geodetic effect does not
appear in the output $S_{\eta_0}$, in other words it produces a purely
$S_{\rho_0}$, or NS signal.  There is an additional contribution to
this precession arising from the Earth's oblateness at about 7 mas/yr,
which must be taken into account 
\cite{wilkins,barker70,oconnell,barkeroconnell3,breakwell,adler}  

The final term in Eq. (\ref{Hdot}) is the Lense-Thirring or
frame-dragging effect.   It has both a
constant term and a periodic term at twice the orbital frequency, 
but for a circular polar orbit, it
produces an orbit-averaged constant rate of 
${\bf J}_\oplus \times {\bf {\bar \xi}} /2R^3 
= J_\oplus \cos \delta /2R^3 \sim 41 \cos \delta$
mas/yr, where $\delta$ is the declination of the guide star.  If the
plane of the polar orbit is oriented so that the guide star location
is in the orbital plane, then the signal is purely $S_{\eta_0}$, or EW. 

The next term in Eqs. (\ref{finalanswer}) is the deflection of light.
For the ecliptic latitude of the guide star, this gives a maximum
deflection caused by the Sun of around 20
mas, once in the one-year mission.  This effect can be included as a
term to be estimated in the data analysis -- studies by the GP-B team
indicate that it could be measured to about one percent -- or it can
calculated {\it a priori} using general relativity and the known
parameters of the Earth's orbit and the guide star location.   
Because the guide star will be
occulted by the Earth, the Earth will also contribute a once-per-orbit
deflection of
around 0.5 mas (which will compete with deflection caused by
propagation of the signal through the Earth's atmosphere).  
Because it averages to zero, this effect can be ignored. 

These relativistic terms, geodetic precession, frame dragging and
deflection of light, have been calculated using the specific
assumptions and constraints inherent in the PPN framework.  An
alternative, more phenomenological viewpoint, would be to regard each as an
independent effect with a separate phenomenological parameter to be
measured in the data analysis.  

The remaining terms in Eqs. (\ref{finalanswer}) are of order $\delta$.
The parallax term caused by the Earth's motion (around
10 mas) must
be taken into account, while that due to the satellite's orbit is
negligible.  Finally the pointing error terms will be controlled to
better than 20 mas (rms), and in any event will be measured directly
by the telescope; all that is needed then is a sufficiently accurate
calibration (via deliberate dithering of the spacecraft axis by known
amounts) to convert telescope readout signals to the appropriate
contributions ${\cal T}_{\eta_0}$ and ${\cal T}_{\rho_0}$
to $S_{\eta_0}$ and $S_{\rho_0}$.  

\section{Conclusions}

We have carried out a covariant calculation of the measurable output
of an orbiting gyroscope experiment, and shown that the result may be
expressed in terms of contributions that can be calculated consistently
using an
Earth-based coordinate frame.  The dominant contributions that are
detectable within the stated 0.4 mas/yr accuracy of the GB-P mission
are: the aberration of starlight and its special relativistic
correction, the general relativistic geodetic precession (due both to
the Earth and to the Sun) and the general relativistic frame-dragging.
The results are in agreement with other, non-covariant calculations,
including numerous unpublished calculations and simulations
carried out by the GP-B team
in the course of
developing the data analysis sytem for the mission.
Because all measurements are referred to instruments fixed on a rolling
spacecraft, the measurable relativistic effects enter mathematically
via their effect
on the apparent direction of the light from the guide star.

\begin{acknowledgments}

We are grateful to Mac Keiser, Francis Everitt and Alex Silbergleit 
for useful discussions, and for important insights into the nature of
the GP-B experiment, and to Charles Pellerin for suggesting the
calculation.
This work was supported in part by the National Science Foundation under
grant No. PHY 00-96522.  

\end{acknowledgments}

\appendix*
\section{Effect of aberration on measurement of roll phase}

The actual signals measured on the GP-B spacecraft are $S_{\eta}$ and
$S_{\rho}$; these are then converted to roll-fixed signals in the data
analysis by forming the combination
\begin{eqnarray}
S_{\eta_0} &=& S_{\eta} \cos \theta - S_{\rho} \sin \theta \,,
\nonumber \\
S_{\rho_0} &=& S_{\eta} \sin \theta + S_{\rho} \cos \theta \,,
\end{eqnarray}
where $\theta$ is the spacecraft roll phase.  In practice this is
determined using a pair of star-trackers fixed to the spacecraft
platform directed roughly perpendicular to the roll axis, 
that continually compare the spacecraft orientation to a
library of known stellar positions, and provide a phase accurate to
a few arcseconds.  The angle $\theta$ is the accumulated phase in the
onboard rotating frame since the
initial state, so that any error in measuring $\theta$ relative to its
``true'' value will generate corrections to the roll-fixed outputs
given by  $\Delta S_{\eta_0} \approx -S_{\rho_0} \Delta \theta$, and 
$\Delta S_{\rho_0} \approx S_{\eta_0}\Delta \theta$.  

Consider a star-tracker whose optical axis is parallel to the 
$\fourvec{\eta}$ vector.  It will record a star when the incoming
projected world line of the photon is anti-parallel to $\fourvec{\eta}$.
i.e. when $\fourvec{\eta}^\prime =- \fourvec{\hat \lambda} = -{\bf n} + {\bf
v} - {\bf n} {\bf n} \cdot {\bf v}$.  In the absence of spacecraft
motion, we would have  $\fourvec{\eta} = - {\bf n}$.  If
$\theta^\prime$ is the measured roll phase and $\theta$ is the
``true'' roll-phase (without aberration), then 
\begin{eqnarray}  
\cos \theta^\prime &=& \fourvec{\eta}^\prime \cdot \fourvec{\eta}_0
\nonumber \\
&=& \cos \theta + {\bf v} \cdot {\bf \rho} \sin \theta \,,
\end{eqnarray}
hence $\Delta \theta = {\bf v} \cdot {\bf \rho} = {\bf v} \cdot {\bf
\rho}_0 \cos \theta - {\bf v} \cdot {\bf \eta}_0 \sin \theta$.  But
since $S_{\eta_0} = S_{\eta_0}(0) - \epsilon^{1/2} 
{\bf v} \cdot {\bf {\bar \eta}}_0 + O(\epsilon,\delta)$, with a
parallel expression for $S_{\rho_0}$, the product $S_{\eta_0} \Delta
\theta$ is of magnitude at most $10^{-8} \sim 2$ mas, but most importantly is
periodic at the roll frequency, and averages to zero.  Aberration will be
taken into account in determination of roll phase from the star trackers.


\begin{thebibliography}{99}
\bibitem{note2} Since $\fourvec{\xi}$ is not strictly an angular
momentum vector, $\fourvec{t}$ is not strictly a torque; in any event,
its precise
form will not be needed.
\end{thebibliography}

\end{document}